\documentclass[10pt,conference]{IEEEtran}
\IEEEoverridecommandlockouts
\usepackage{comment}
\usepackage[sort]{cite} 
\usepackage{amsmath,amssymb,amsfonts}
\usepackage{algorithmic}
\usepackage{graphicx}
\usepackage{subfigure}
\usepackage{textcomp}
\usepackage{xcolor}
\def\BibTeX{{\rm B\kern-.05em{\sc i\kern-.025em b}\kern-.08em
    T\kern-.1667em\lower.7ex\hbox{E}\kern-.125emX}}
\begin{document}

\title{Optimizing RIS Impairments through Semantic Communication \\

\thanks{
This work was partially funded by the French government under the France 2030 ANR program “PEPR Networks of the Future” (ref. 22-PEFT-0010) and by the 6G-GOALS Project under the HORIZON program (no. 101139232). The work of Luca Sanguinetti was supported by the Italian Ministry of Education and Research (MUR) in the framework of the FoReLab project (Departments of Excellence) and the Garden Project (PRIN 2022 program).}
}
\vspace{-16mm}
\author{\fontsize{1}{1}\selectfont
\IEEEauthorblockN{\fontsize{11}{12}\selectfont
Nour Hello\IEEEauthorrefmark{1}, Mattia Merluzzi\IEEEauthorrefmark{1}, Emilio Calvanese Strinati\IEEEauthorrefmark{1}, Luca Sanguinetti\IEEEauthorrefmark{2}
\vspace{0mm}}
\IEEEauthorrefmark{1}CEA-Leti, Université Grenoble Alpes, F-38000 Grenoble, France \\ \IEEEauthorrefmark{2}Dipartimento di Ingegneria dell’Informazione, University of Pisa, Pisa, Italy \\ \fontsize{8}{8}Emails: \{nour.hello, mattia.merluzzi, emilio.calvanese-strinati\}@cea.fr, luca.sanguinetti@unipi.it
\vspace{-2mm}
}

\maketitle

\begin{abstract}
This paper investigates how semantic communication can effectively influence and potentially redefine the limitations imposed by physical layer settings. 
Reconfigurable Intelligent Surfaces (RIS) enable the intelligent configuration of the physical layer of communication systems. However, its practical implementation is hampered by several limitations.
The Semantic Communication (SemCom) paradigm introduces additional degrees of freedom that can be exploited to improve the robustness of communication against physical layer impairments.
In essence, SemCom ensures that the data representation remains robust even under adverse physical conditions by emphasizing the transmission of meaningful information in a manner that is less susceptible to degradation. Through the use of SemCom, potential RIS gains are demonstrated in terms of RIS area size and the phase shift precision of its active elements.
\end{abstract}


\section{Introduction}
The transition from 5G to 6G will mark a profound transformation in how networks and technologies interact with humans across physical and cyber spaces \cite{Uusitalo2021}. This will create intelligent and programmable physical and digital environments that will provide ubiquitous wireless intelligence as a service \cite{Merluzzi2023Hex}.
Multiple novel technology breakthroughs underpin 6G, including reconfigurable intelligent surface and semantic communication.  

First, the new concept of \textit{programmable wireless environment} aroused, capitalizing on research, design, and prototyping of new network elements called Reconfigurable Intelligent Surfaces (RISs). 
The idea is that with RISs, which are (large) surfaces composed of nearly passive reconfigurable elements, the radio wave propagation can be intentionally controlled to realize desired dynamic transformations on the wireless propagation environment, namely reflections that do not necessarily obey Snell's law. In RIS-assisted systems, each element receives the signal from the transmitter and modifies its phase before reflecting or scattering it to the destination \cite{Mukherjee2021}. This dynamic and goal-oriented control over radio wave propagation introduces the concept of "\textit{wireless environment as a service}" \cite{CalvaneseRISE-6G2021}.
The gain of an RIS in shaping the propagation channel and enhancing network performance depends on two main factors: the surface size and structure, and the precision of phase shifts, mainly related to hardware realizations and constraints. Larger surfaces capture and redirect more energy, resulting in higher end-to-end channel power gain, which theoretically scales with the square of the number of elements $(N_r^2)$, assuming optimal phase alignment and negligible mutual coupling effects. It should be noted that this gain should compensate for the inevitable additive path loss effects, which dramatically decrease the received power in an RIS-aided communication link. Additionally, phase shift precision affects the overall gain. This precision depends on two key aspects: phase quantization precision and freshness of Channel State Information (CSI). In practical RIS implementations, phase shifts are often quantized (e.g., with 1 to 3 bits in recent implementations), meaning they can only take on discrete values. Higher quantization precision allows for finer adjustments, leading to more accurate phase alignment of the reflected waves. Additionally, the phase shift applied by each element should be based on accurate and up-to-date CSI, which can be expensive to acquire in the case of large surfaces, in terms of time-frequency-energy resources. If the CSI is outdated or partial, the applied phase shifts may not align correctly with the current channel conditions, resulting in sub-optimal performance and reduced gain.

Second, a fundamental pillar of 6G networks involves embedding artificial intelligence (AI) techniques directly into the design and operation of wireless communication systems, known as \textit{AI-native} systems. Unlike traditional methods that add AI as an ad-hoc optimization layer, AI-native systems integrate AI throughout the entire protocol stack and air interface. Recent research is showing the notable benefits offered by \textit{Semantic communications} (SemComs) to support relevant, effective, and timely exchange of messages between intelligent agents capable of extracting and understanding the underlying meaning (semantic) within a data stream \cite{CalvaneseGOWSC2021}. This approach offers enhanced robustness against wireless noise, interference, and fading by focusing on encoding and decoding the meaning of the transmitted data in contrast to classical methods of message construction and recovery. The SemCom offers new compression-robustness trade-offs, enabling to further reduce the use of bandwidth, latency, and energy resources, thus questioning the need for ultra-high communication speed and providing a robust solution to \textit{transmit less to achieve more}. Broadly, the SemCom paradigm introduces numerous additional degrees of freedom that can be effectively leveraged to enhance network efficiency and improve communication robustness against physical layer impairments. In this paper, we investigate the joint optimization of RIS and semantic communication systems, focusing on leveraging SemComs to reduce the complexity associated with RIS. We focus on the semantic encoding of knowledge graphs (KGs) to serve as a universal representation of multimodal data. The embeddings of KGs in a latent space are modulated as semantic symbols for transmission. Subsequently, the receiver decodes and infers the KG structure and attributes.

\textbf{Related works.} In the recent literature, \cite{Shi2023} introduces a semantic communication model for text transmission that utilizes RIS to enhance channel quality in point-to-point communication systems. \cite{Jiang2023} leverages the application of RIS to semantic image transmission, using reinforcement learning to optimize RIS elements and semantic features codeword allocation based on estimated channel conditions and user requirements. In multi-user systems, \cite{Jiang2024} demonstrates efficient allocation of RIS resources with a focus on user requirements priorities. \cite{Zhao2024} combines distributed RISs with probabilistic semantic communication, addressing a semantic-aware sum rate maximization problem that includes constraints on transmit power, RIS-user association, phase shift, and semantic compression ratio. Additionally, \cite{Ma2024} proposes a hierarchical optimization framework for RIS-assisted semantic communication systems, where the RIS module is first optimized to create desirable physical channel conditions, and then the neural networks in the semantic layer are trained end-to-end. This proposed system by \cite{Ma2024} exhibits greater robustness in low-to-medium Signal-to-Noise Ratio (SNR) regimes compared to end-to-end semantic communication systems without RIS.
\\ \textbf{Contribution.} 
In this paper, we harness semantic communication to relax stringent physical layer constraints, thereby enhancing flexibility, efficiency, and performance in practical implementations. By employing a semantic communication system grounded in knowledge graph representation learning, fewer reflecting elements of the RIS are needed to infer the knowledge graph and subsequently its equivalent text, compared to traditional encoding methods. This reduction in required elements potentially lowers overall power consumption, channel estimation costs, etc., thereby preserving energy efficiency. Furthermore, the semantic approach necessitates fewer bits for the quantization of phase beamforming of the RIS reflecting elements. Therefore, this paper takes the complementary perspective on RIS-aided communications, with SemComs representing a means for overcoming RIS limitations.
\section{Overall System Model}
\begin{figure*}[htbp]
\centering
\includegraphics[width=.8\textwidth]{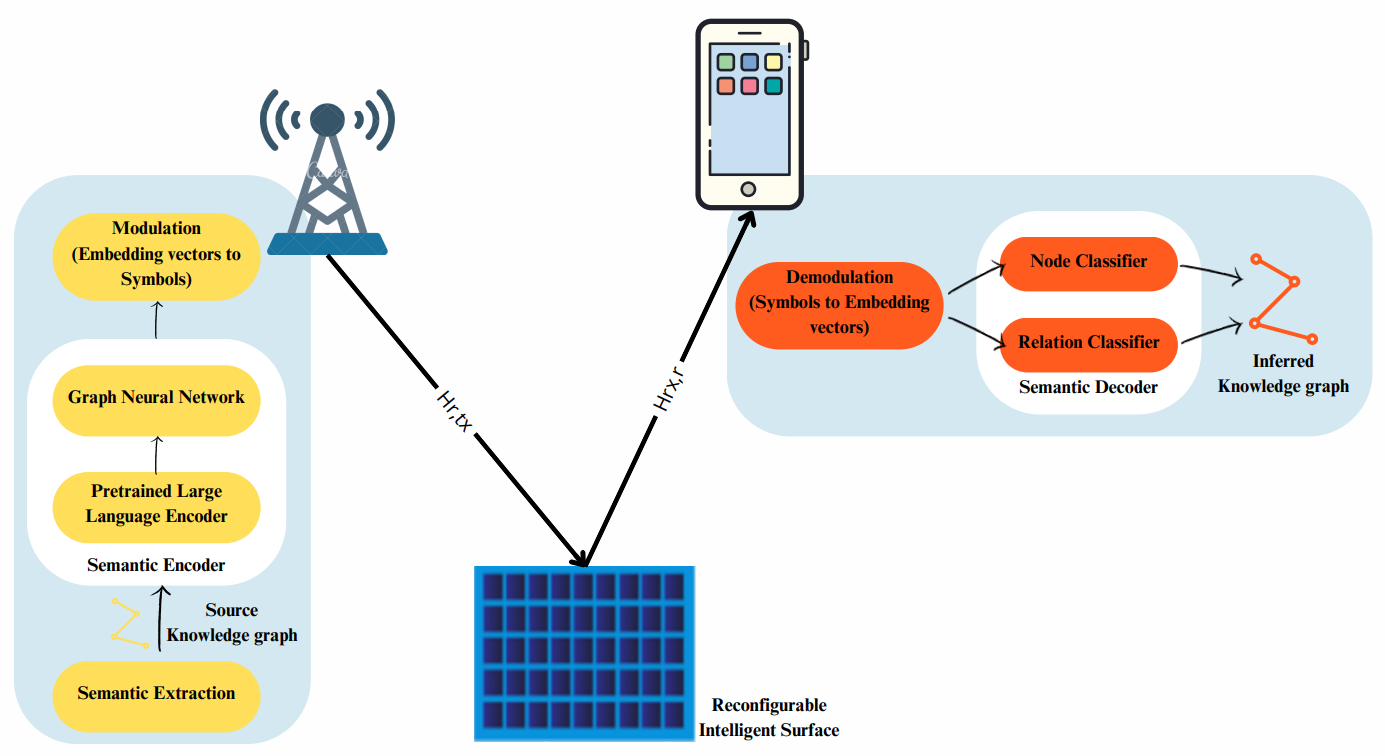}
\caption{System model of semantic representation learning on graphs via reconfigurable intelligent surfaces}
\label{fig:diag}
\end{figure*}
SemComs establish a dual communication channel comprising semantic and wireless components.
Notably, the semantic channel involves entailment relations between "models" and "messages" within the source \cite{BASU2014}.
The overall impact of both the semantic channel and the wireless channel on the input data, 
leading to the output of a semantic decoder, can be expressed mathematically as a composition of transformations applied over the semantic space. These transformations capture how the data is modified as it passes through both channels, ultimately influencing the output produced by the semantic decoder. Therefore, We can express the overall transformation introduced by the E2E semantic system $\mathbf{T}_\text{e2e, sc}$ as follows:
\begin{equation}
\begin{split}
\mathbf{T}_\text{e2e, sc}(\textit{X}) = Ts_1 \circ \ldots \circ Ts_j  \circ Tw_1 \circ  \ldots \circ Tw_{k} \\
 \circ To_1 \circ \ldots \circ To_n (\textit{X})
\end{split}
\end{equation}
Here, $Ts_i$ represents a transformation operating at the semantic channel, $Tw_i$ represents a transformation operating at the wireless channel, and $To_i$ represents a transformation 
operating at other modules of the semantic channel, such as semantic channel equalization modules \cite{sana2023}.
The performance and characteristics of each transformation are influenced by several key factors that collectively shape the effectiveness and practicality of applying it in various contexts. These factors include the communication environment, available side information, composition with other transformations, estimation cost, system architecture, and system reconfiguration costs.
This paper focuses on exploring the disjoint effect of wireless channels shaped through RISs within the context of communicating the semantic representation of knowledge graphs.
In the subsequent subsections, we expand on the characteristics of the semantic transformations entailed on knowledge graph data and the wireless transformations induced by the RIS, pictured by our system model in  Fig.~\ref{fig:diag}.

\subsection{Graph-based Semantic communication}\label{sec:semantic}
One key aspect of SemCom entails compressing knowledge representation. This aspect is often confined to exploiting Deep Learning (DL)-based Joint Source-Channel Coding. Specifically, the transmitter leverages background knowledge relevant to source messages to filter out irrelevant content and extract core features that require fewer bits for transmission \cite{CalvaneseGOWSC2021}. 
We follow the semantic system introduced in\cite{Hello2024}. In this end-to-end semantic system, a source KG is encoded into a set of embedding vectors in a latent space, such that the number of vectors equals the number of nodes in the graph. Subsequently, each vector is modulated as complex symbols that are transmitted over the wireless channel. Then, the receiver demodulates the received noisy symbols to vectors and, eventually, infers the knowledge graph.
The KG is defined as a set of interconnected triplets, each triplet is organized in the form of (node-relation-node). Mathematically, a KG is denoted as $\mathcal{G} = (\mathcal{N}, \mathcal{E})$, where $\mathcal{N} = \{\textit{n}_i\}_{i=1}^{\mathbf{N}_e}$ represents the set of nodes, and $\mathcal{E} = \{\textit{r}_{ij}\}_{i,j=1}^{\mathbf{N}_e}$ is the set of relational edges associating these nodes. $\textit{n}_i$ denotes the textual attribute associated with a node of index $i$, $\textit{r}_{ij}$ denotes the textual attribute of the link from a source node of index $i$ to a target node of index $j$, while the parameter $\mathbf{N}_e$ specifies the total number of nodes within the graph.
\subsubsection{Semantic Encoder}
 The semantic encoder architecture is a sequence of a pretrained large language model (LLM) encoder and a Graph Neural Network (GNN) that represents the input knowledge graph $ \mathcal{G}$ in semantic latent space as a matrix of low-dimensional vectors expressed as $\mathbf{X}^{\prime}$, denoted in \eqref{Xprime}, such that each row of $\mathbf{X}^{\prime}$ is a vector \( \mathbf{x}^{\prime}_i \) associated with a node of index $i$. Extensively, the pretrained LLM generates the embedding vectors of the textual attributes associated with nodes and relations within $\mathcal{G}$ as delineated in \eqref{llm_eq}. Afterward, as outlined in \eqref{gnn_eq}, the GNN compresses those embedding vectors into the matrix $\mathbf{X}^{\prime}$ with the number of rows equals to the number of nodes in $\mathcal{G}$, and the number of columns equals to the embedding dimension.

   \begin{equation}\label{Xprime}
    \mathbf{X}^{\prime} = 
\begin{bmatrix}
\mathbf{x}^{\prime}_1 \\
\mathbf{x}^{\prime}_2 \\
\vdots \\
\mathbf{x}^{\prime}_{\mathbf{N}_e}
\end{bmatrix}.
\end{equation}

\begin{equation} \label{llm_eq}
    \mathbf{X} = \mathbf{LLM}(\mathcal{N}), \hspace{0.5cm}
    \mathbf{E}= \mathbf{LLM}(\mathcal{E}),
\end{equation}
\begin{equation} \label{gnn_eq}
    \mathbf{X}^{\prime} =\mathbf{GNN}(\mathbf{X}, \mathbf{E}).
\end{equation}

\subsubsection{Channel Encoding}
The channel encoder block involves a feed-forward neural network (FNN) followed by a power normalization layer. The channel encoder modulates each low-dimensional vector $\mathbf{x}^{\prime}_i$ encoded by the semantic encoder as $n$ complex symbols. This can be represented mathematically as
\begin{equation} \label{fnn_eq}
    \mathbf{S} =\mathbf{FNN}(\mathbf{X}^{\prime});
\end{equation}
where $\mathbf{S}$ is denoted as:
\begin{equation}
    \mathbf{S} = 
\begin{bmatrix}
s_{1,1} & s_{1,2} & \cdots & s_{1,n} \\
s_{2,1} & s_{2,2} & \cdots & s_{2,n} \\
\vdots & \vdots & \ddots & \vdots \\
s_{N_e,1} & s_{N_e,2} & \cdots & s_{N_e,n}
\end{bmatrix}.
\end{equation} 
Each row of $\mathbf{S}$ is associated with a given node of the graph $\mathcal{G}$, and is a row of $n$ complex symbols, one complex symbol is designated in the following as $s_{ij}$. The power normalization ensures that the energy of the $n$ symbols constellation averages to $1$. Upon reception, the channel decoder decodes the matrix of received symbols $\mathbf{\hat S}$ into the semantic representation latent space matrix denoted as $\mathbf{\hat X}^{\prime}$, through an FNN architecture.
\subsubsection{Semantic Decoder}
Two sub-modules of the semantic decoder are combined to infer the knowledge graph $\mathcal{\hat G}$ from the low-dimensional latent space matrix $\mathbf{\hat X}^{\prime}$ denoted as:
\begin{equation}
    \mathbf{\hat X}^{\prime} = 
\begin{bmatrix}
\mathbf{\hat x}^{\prime}_1 \\
\mathbf{\hat x}^{\prime}_2 \\
\vdots \\
\mathbf{\hat x}^{\prime}_{\mathbf{N}_e}
\end{bmatrix};
\end{equation}
where each row of $\mathbf{\hat X}^{\prime}$ is a vector \( \mathbf{\hat x}^{\prime}_i \) associated with a node of index $i$. The node classification mechanism modeled as a multi-layer perceptron with skip connections classifies $\mathbf{N}_e$ distinct low-dimensional $\mathbf{\hat x}^{\prime}_i$ vectors, assigning each of them to a specific node descriptor $\tilde{\textit{n}}_i$.
In conjunction, the relation decoder, comprising a vanilla transformer encoder with a classification head, discerns the adjacency matrix of $\mathcal{\hat G}$ from the latent space matrix $\mathbf{\hat X}^{\prime}$, along with the textual attributes associated with the inferred edges.

\subsection{RIS-aided communications}

RIS improve wireless connectivity by shaping propagation paths, even when line-of-sight (LoS) links are blocked. A passive RIS reflects signals in intended directions, providing beamforming gain to mitigate multiplicative path loss. While this gain is theoretically proportional to $N_r^2$ (with $N_r$ being the number of RIS elements), practical implementations face challenges such as phase quantization, hardware imperfections, and channel estimation errors. 
RIS-aided communications involves the MIMO channel $\mathbf{H}_{r,\text{tx}}\in\mathbb{C}^{N_r\times N_{\text{tx}}}$ between the transmitter and the RIS, and the MIMO channel $\mathbf{H}_{\text{rx},r}\in\mathbb{C}^{N_{\text{rx}}\times N_r}$ between the RIS and the receiver. Both channels typically include a LoS path and an NLoS path. In this paper, we assume direct communication with no NLoS component, and we hypothesize that the direct link between the transmitter and receiver is absent. 
The $(m,n)$-th element of $\mathbf{H}_{i,j}$ between transmitter $j$ and receiver $i$ reads as \cite{Demmer2023}:
\begin{equation}\label{channel_tx_rx}
    {H}_{i,j}^{(m,n)} = \sqrt{\pi^2\cos{(\varphi^{i,j}_{mn})}\cos{(\varphi^{j,i}_{mn})}\beta_{i,j}^{-1}}e^{-j\kappa d_{i,j}^{(m,n)}},
\end{equation}
where $d_{i,j}^{(m,n)}$ denotes the distance between the antenna elements, $\beta_{i,j}$ denotes the path loss between transmitter $j$ and receiver $i$, and $\kappa=\frac{2\pi}{\lambda}$ is the wave number, with $\lambda$ being the wavelength. Also, $\varphi^{i,j}_{mn}$ is the azimuth angle between the broadside direction of array $i$ and array $j$. The $\cos(\cdot)$ terms represent the non-isotropic response profile of the array elements \cite{Demmer2023}. Exploiting \eqref{channel_tx_rx} for the two channels tx $\rightarrow r$ and $r\rightarrow$ rx, and assuming that no direct LoS path is available between tx and rx (as also illustrated schematically in Fig. \ref{fig:diag}, the end-to-end channel reads as:
\begin{equation}\label{e2e_channel}
\mathbf{H}_{\text{e2e}}=\mathbf{H}_{r,\text{tx}}\mathcal{Q}(\mathbf{\Theta})\mathbf{H}_{\text{rx},r},
\end{equation}
where $\mathbf{\Theta}$ is a diagonal matrix, whose $i$-th diagonal element reads as $e^{j\theta_i}$, with $\theta_i\in[0,2\pi]$ the \textit{continuous} phase shift of the $i$-th RIS element; whereas, $\mathcal{Q}(\cdot)$ denotes the non-linear transformation that comes from phase quantization. Given a modulated semantic symbol expressed as $s_{ij}$  (as described in section \ref{sec:semantic}), the received semantic symbol through the RIS channel after combining at the receiver  is:
\begin{equation}\label{rx_symbol}
\hat s_{ij}=\mathbf{w}_{\text{rx}}^H\mathbf{H}_{r,\text{tx}}\mathcal{Q}(\mathbf{\Theta})\mathbf{H}_{\text{rx},r}\mathbf{w}_{\text{tx}}\sqrt{P_{\text{tx}}}s_{ij} + n,
\end{equation}
where $\mathbf{w}_{\text{tx(rx)}}$ denotes the transmitter precoding (receiver combining), $n$ the noise at the receiver, and $P_{\text{tx}}$ the transmit power.
It's worth noting that the channel remains constant through out the transmission of the matrix of semantic symbols $\mathbf{S}$.
Then, the SNR at the receiver can be written as:
\begin{equation}\label{SNR}
    \text{SNR} = \frac{||\mathbf{w}_{\text{rx}}^H\mathbf{H}_{r,\text{tx}}\mathcal{Q}(\mathbf{\Theta})\mathbf{H}_{\text{rx},r}\mathbf{w}_{\text{tx}}||^2 P_{\text{tx}}}{\sigma^2}.
\end{equation}
This SNR remains the same for the transmission of each semantic symbol in $\mathbf{S}$.
\section{Evaluation Study}
\begin{figure*}[htbp]
    \centering
    \subfigure[One bit phase quantization]{
        \includegraphics[width=0.3\textwidth]{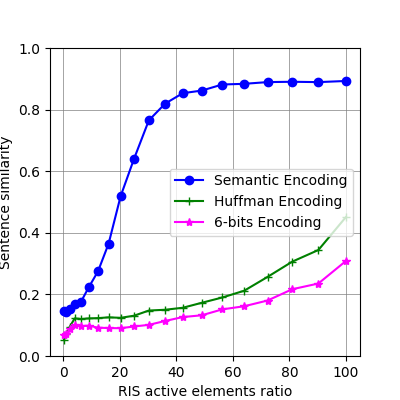}
        \label{fig:one_sbert}
    }
    \hfil
    \subfigure[Two bits phase quantization]{
        \includegraphics[width=0.3\textwidth]{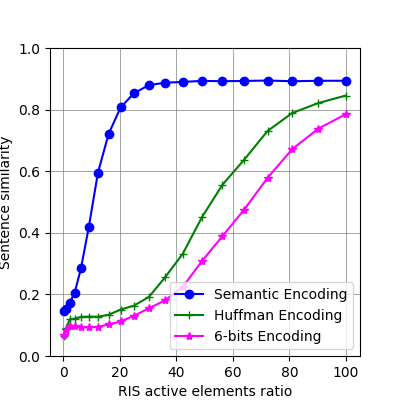}
        \label{fig:two_sbert}
    }
    \hfil
    \subfigure[No phase quantization]{
        \includegraphics[width=0.3\textwidth]{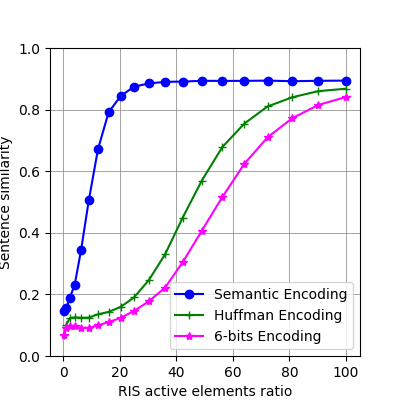}
        \label{fig:no_sbert}
    }
    \caption{Sentence similarity versus RIS active elements ratio}
    \label{fig:sbert}
\end{figure*}
\begin{figure*}[htbp]
    \centering
    \subfigure[One bit phase quantization]{
        \includegraphics[width=0.3\textwidth]{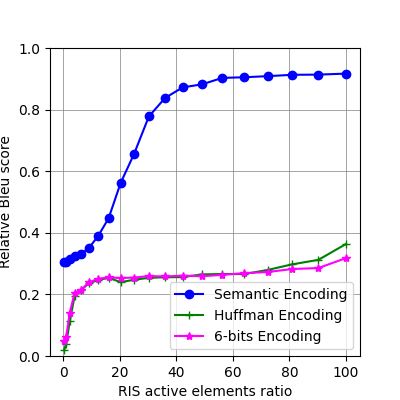}
        \label{fig:o_bleur}
    }
    \hfil
    \subfigure[Two bits phase quantization]{
        \includegraphics[width=0.3\textwidth]{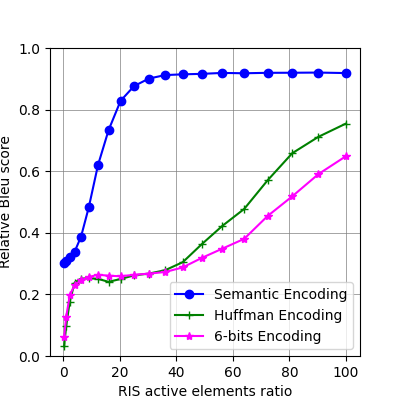}
        \label{fig:t_bleur}
    }
    \hfil
    \subfigure[No phase quantization]{
        \includegraphics[width=0.3\textwidth]{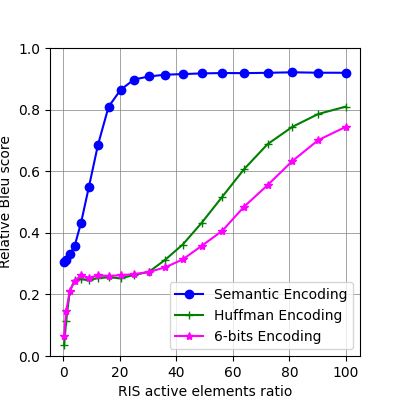}
        \label{fig:n_bleur}
    }
    \caption{Relative Bleu Score versus RIS active elements ratio}
    \label{fig:bleur}
\end{figure*}
\begin{figure*}[htbp]
    \centering
    \subfigure[One bit phase quantization]{
        \includegraphics[width=0.3\textwidth]{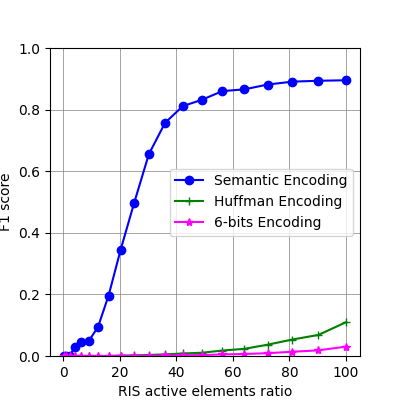}
        \label{fig:one_f1}
    }
    \hfil
    \subfigure[Two bits phase quantization]{
        \includegraphics[width=0.3\textwidth]{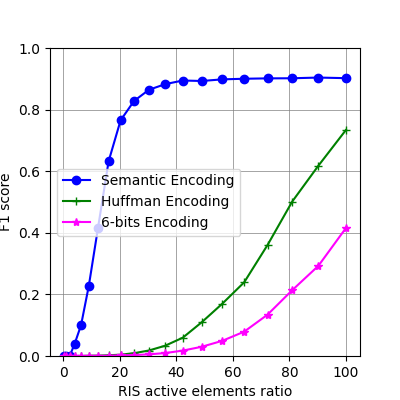}
        \label{fig:two_f1}
    }
    \hfil
    \subfigure[No phase quantization]{
        \includegraphics[width=0.3\textwidth]{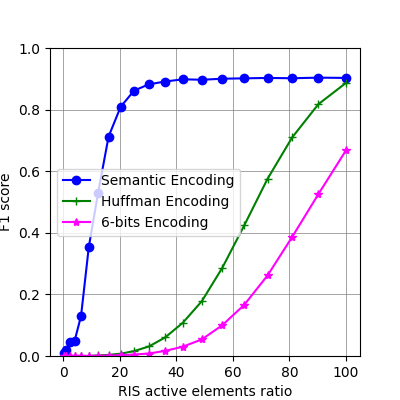}
        \label{fig:no_f1}
    }
    \caption{F1 score of knowledge graph matching versus RIS active elements ratio}
    \label{fig:f1}
\end{figure*}
This section presents numerical results to evaluate the gains offered by the semantic model based on knowledge graph representation learning (section \ref{sec:semantic}) over traditional encoding methods (Huffman and 6-bits) in RIS communication scenarios, assuming the absence of a direct LoS path between the transmitter and receiver.
\subsection{System parameters}
We simulate a two-stage system that leverages the semantic model and the RIS to optimize the wireless environment.
\subsubsection{Communication-related parameters}
We consider a scenario with the transmitter, receiver, and RIS are placed at 3D coordinates $[0,10,0]$, $[10,15,0]$, and $[10,0,0]$, respectively. The transmitter is equipped with a $10\times 10$ uniform planar array (UPA), with antenna element spacing equal to $\lambda/2$ and the transmit power is set to $0.1$ W. The carrier frequency is $28$ GHz. The transmitter precoding is optimized using antenna theory to maximize gain in the direction of the RIS \cite{Clemente2012}. The RIS is also a UPA with element spacing $\lambda/2$. For the different simulations, we will refer to the number of \textit{active} RIS elements, as the number of RIS elements that participate in the reflection. Equivalently, this would mean considering a more or less large RIS. The largest RIS we consider (i.e., $100$\% of RIS active elements) is a $40\times40$ UPA. This predefined codebook is built as in \cite{airod2023}, using antenna array theory as in \cite{Clemente2012}. The path loss exponent is $4$, and the noise power at the single antenna receiver is set to $-120$ dBm.
\subsubsection{Semantic level setup}
At the semantic encoder side, we set up the pre-trained language model encoder ‘all-MiniLM-L12-v2’ \cite{reimers-2019-sentence-bert}, which, given a knowledge graph, associates feature vectors with the descriptors of its nodes and edges, each of size 384. 
As in \cite{Hello2024}, we instantiate the GNN as the graph isomorphism convolutional neural network \cite{gine}. The GNN exploits the relational structure of the graph to compress the feature vectors of its nodes to latent space vectors, each of size 16. Subsequently, each embedding vector of size 16 is modulated as 3 complex symbols. The end-to-end semantic model is trained on the WebNLG dataset \cite{webnlg} at an SNR of 14 dB, with a batch size of 8 graphs. We also fine-tuned the T5 encoder-decoder transformer on the WebNLG dataset to generate text corresponding to a noiseless knowledge graph separately from the training of the end-to-end semantic model.
The T5 transformer achieves a sentence similarity score $\mathbf{max sim}=0.93$ and a Bleu (Bilingual Evaluation Understudy) score $\mathbf{max bleu}=0.6$ between the text generated from the source noiseless knowledge graph and the source text.
\subsection{Methodology}
At the testing stage, we assume that the RIS is configured using a codebook generated as described in \cite{airod2023}, applying antenna array theory from \cite{Clemente2012}. For continuous phases, the selected codebook index is the one that generates the highest SNR, based on the models in \eqref{e2e_channel}, \eqref{rx_symbol}, and \eqref{SNR}. Then, the phases are quantized using uniform quantization with resolutions of 1 and 2 bits.
Subsequently, the trained semantic model's performance is tested on the different SNR values depending on the RIS active elements ratio, and the phase shift precision. The RIS active elements ratio refers to the percentage of RIS elements employed in the communication between tx and rx, with the full RIS being a $40\times 40$ UPA.
\subsection{Numerical Results}
Fig.~\ref{fig:sbert}  shows semantic textual similarity versus RIS active elements ratio for one (Fig.~\ref{fig:one_sbert}), two bits (Fig.~\ref{fig:two_sbert}) and no phase quantization (Fig.~\ref{fig:no_sbert}).
To compute Semantic Textual Similarity, we generate, based on \cite{reimers-2019-sentence-bert}, embeddings for both the text inferred from the decoded noisy knowledge graph and the original text. We then calculate the similarities between these embeddings using the cosine similarity metric.
Fig.~\ref{fig:bleur} shows the ratio of the Bleu score over $\mathbf{max bleu}$ versus RIS active elements ratio for one (Fig.~\ref{fig:o_bleur}), two bits (Fig.~\ref{fig:t_bleur}) and no phase quantization (Fig.~\ref{fig:n_bleur}).
The Bleu score is limited to measuring token overlap between inferred texts and reference texts, rather than assessing the actual meaning of the text.
Both Fig.~\ref{fig:one_sbert} and Fig.~\ref{fig:o_bleur} show that, given a phase quantization precision of one-bit, the proposed semantic method achieves its optimal performance for $55\%$ active RIS elements, while the traditional encoding methods fail to achieve an acceptable performance for $100\%$ active elements.\\
Fig.~\ref{fig:two_sbert} and Fig.~\ref{fig:t_bleur} show that the semantic model requires $20\%$ less active elements to achieve its optimal performance for two bits over one bit phase quantization. 
Fig.~\ref{fig:no_sbert} and Fig.~\ref{fig:n_bleur} show that the semantic encoding scheme needs only $20\%$ RIS active elements to reach its optimal performance for no phase quantization. In those settings, the Huffman encoding approach slightly achieves this optimal performance for $100\%$ RIS active elements in terms of the semantic textual similarity, while the Bleu score is $14.5\%$ higher for the semantic approach compared to the Huffman encoding method for $100\%$ RIS active elements.\\
 At the knowledge graph data level, Fig.~\ref{fig:f1} shows the F1 score measuring the accuracy of graphs inference versus RIS active elements ratio for one (Fig.~\ref{fig:one_f1}), two bits (Fig.~\ref{fig:two_f1}) and no phase quantization (Fig.~\ref{fig:no_f1}).
 The F1 score evaluates the accuracy of triplets inference by assessing the exact correspondence between elements in the source and decoded knowledge graphs.
 It measures the harmonic mean of precision and recall, reflecting how well the inferred triplet elements match the source triplet elements.
 Fig.~\ref{fig:f1} validates the findings presented in Fig.~\ref{fig:sbert} and Fig.~\ref{fig:bleur} while extending the analysis beyond text applications:
It assesses the gains brought by the semantic method over traditional encoding approaches in the RIS wireless settings for the broader data structure of knowledge graphs, which serve as a universal representation of data. This validation underscores the applicability of the semantic method across diverse data types beyond text in optimizing the gain of the RIS depending on its number of active elements and the precision of the phase shifts of its elements.

\section{Conclusions}
The disjoint optimization of semantic and RIS components improves the reliability and efficiency of data transmission, allowing the RIS gain to be optimized in terms of the number of active elements and the precision of their phase shifts. The integration of semantic communication with RIS technology promises to overcome inherent implementation challenges and optimize performance in next-generation wireless networks. In addition, this approach demonstrates scalability by enabling RIS settings to support more users without requiring additional physical resources and it can facilitate the coexistence and interoperability of semantic and non-semantic users.
\bibliographystyle{IEEEtran} 
\bibliography{References} 

\end{document}